\documentclass[a4paper,12pt]{article}

\usepackage[pdftex]{graphicx}
\usepackage[colorlinks]{hyperref}

\providecommand{\keywords}[1]
{
  \small
  \textbf{Keywords:} #1
}

\usepackage{amsmath,amssymb,amsthm}
\usepackage{bm} 
\usepackage{mathtools}
\usepackage{booktabs}
\usepackage{cite}

\theoremstyle{definition}
\newtheorem{example}{Example}

\makeatletter
    
    \@addtoreset{equation}{section}
\makeatother

\title{\Large\textbf{Harmonic Oscillator with a Step and its Isospectral Properties}}
\author{\large Yuta Nasuda\thanks{Corresponding author. E-mail: \texttt{y.nasuda.phys@gmail.com}, ORCiD: \texttt{0000-0002-0149-8483}} ~~
	and ~
	Nobuyuki Sawado\thanks{E-mail: \texttt{sawadoph@rs.tus.ac.jp}, ORCiD: \texttt{0000-0002-9740-7792}} \\[1ex] 
	{\normalsize Department of Physics and Astronomy, Tokyo University of Science, Noda,} \\
	{\normalsize Chiba 278-8510, Japan}}
\date{\footnotesize (Dated: \today)}

\allowdisplaybreaks[4]

\begin{document}

\maketitle

\begin{abstract}
We investigate the one-dimensional Schr\"{o}dinger equation for a harmonic oscillator with a finite jump $a$ at the origin.
The solution is constructed by employing the ordinary matching-of-wavefunctions technique.
For the special choices of $a$, $a=4\ell$ ($\ell=1,2,\ldots$), the wavefunctions can be expressed by the Hermite polynomials.
Moreover, we explore isospectral deformations of the potential via the Darboux transformation.
In this context, infinitely many isospectral Hamiltonians to the ordinary harmonic oscillator are obtained.
\end{abstract}

\keywords{
exactly solvable models, Schr\"{o}dinger equation, matching of wavefunctions, piecewise analytic functions, Hermite polynomials, isospectral Hamiltonians, Darboux transformation
}

\section{Introduction}
\label{sec:Intro}

Exactly solvable problems in non-relativistic quantum mechanics have been attracting scientists' interest in understanding the behavior of particles in various potential energy landscapes.
More specifically, exact solutions of the time-independent, one-dimensional Schr\"{o}dinger equation:
\begin{equation}
\left[ -\frac{d^2}{dx^2} + V(x) \right] \psi(x) = E\psi(x) ~,
\label{eq:SE}
\end{equation}
have been constructed for a variety of potentials $V(x)$.
The square potential well and the harmonic oscillator are two of the most significant examples that show up in a wide range of fields in quantum physics.

The exactly solvable potentials can be categorized into several classes.
One major class would be the piecewise constant interactions (See, \textit{e.g.}, Ref.\,\cite{Magyari}), including so-called point interaction models~\cite{PointInteraction}.
A typical example is the Kronig--Penney model~\cite{Kronig1931}. 
Another major class is analytical potentials whose eigenfunctions are written in closed analytical form.
The shape-invariant potentials~\cite{Gendenshtein:1983skv} are the typical examples of this class, which are listed in Ref.\,\cite{COOPER1995267} and also found in Refs.\,\cite{Quesne_2008,ODAKE2009414,ODAKE2010173,Sasaki_2010,ODAKE2011164}.
A substantial feature of these examples is that they are solved by the classical orthogonal polynomials or the exceptional/multi-indexed orthogonal polynomials~\cite{GOMEZULLATE2009352,GOMEZULLATE2010987,ODAKE2009414,ODAKE2010173,Sasaki_2010,ODAKE2011164}.
Several deformations of this class of potentials have been considered by employing the Darboux transformation~\cite{darboux}, the Abraham--Moses transformation~\cite{PhysRevA.22.1333} and other techniques (See Fig.\,1 in Ref.\,\cite{10.1007/978-981-19-4751-3_29} and the references therein).

The problems lying in the intersection between these two classes, that is, potentials defined by piecewise analytic functions, have also attracted attention.
For instance, the Coulomb plus square-well potential~\cite{PhysRev.71.865}, a finite parabolic quantum well potential~\cite{PhysRevB.48.17316} and the harmonic oscillator potential embedded in an infinite square-well~\cite{10.1119/1.15738} are discussed in atomic and nuclear physics.
Moreover, the sheared harmonic oscillator~\cite{doi:10.1021/j100356a004}, the inverse square root potential~\cite{Ishkhanyan_2015}, the symmetrized Morse-potential short-range interaction~\cite{doi:10.1142/S0217732316500887}, and other potentials such as $-g^2\exp(-|x|)$-potential~\cite{Sasaki_2016}, a double well potential $\min[(x+d)^2,(x-d)^2]$ and its dual single well potential $\max[(x+d)^2,(x-d)^2]$~\cite{quantum4030022,10.1063/5.0127371} are all defined piecewise and solved by the matching of wavefunctions.

Yet another example exists.
A Dirac delta function at the origin under the harmonic oscillator has been discussed~\cite{Viana-Gomes_2011}.
The energy spectrum of this system is identical to that of the ordinary harmonic oscillator, \textit{i.e.},  they are isospectral, for odd-parity states, while it is not trivial for even-parity states.
Here, for the odd-parity states, the wavefunctions are also the same as the ordinary harmonic oscillator.
This potential can be abstracted as the harmonic oscillator plus a singularity function (For the singularity functions, see, \textit{e.g.}, Ref.\,\cite{lighthill_1958}).

Isospectral deformation of Hamiltonian is of importance, for it is certainly related to the construction of exactly solvable potentials~\cite{darboux,10.1093/qmath/6.1.121,Kre57,Adler:1994aa,PhysRevA.22.1333,PhysRevD.33.1048}.
It has extensively been explored by many authors (For recent works, see, \textit{e.g.}, \cite{PhysRevD.94.105022,Cariena_2017,Guliyev:2020aa,Jafarov:2021aa}).
Some isospectral deformations produce a deformed Hamiltonian that shares many, if not all, eigenvalues with the original Hamiltonian, while other isospectral deformations construct a deformed Hamiltonian whose energy spectrum is entirely identical to the original one's.
The former kind of isospectrality is sometimes referred to as \textit{essential} (or \textit{quasi/partial}) isospectrality, and the later is called \textit{strict} (or \textit{complete}) isospectrality.
For instance, the system with the harmonic oscillator and a Dirac delta function at the origin~\cite{Viana-Gomes_2011} is said to be essentially isospectral to the harmonic oscillator by itself.

In this paper, we discuss the harmonic oscillator with a Heaviside step function, which is a member of the singularity functions:
\begin{equation}
V(x) = \begin{cases}
	\omega^2x^2 - \omega - a & (x<0) \\
	\omega^2x^2 - \omega & (x>0)
\end{cases} ~,
\label{eq:pot_om}
\end{equation}
in which $a$ is an arbitrary constant, and solve the time-independent Schr\"{o}dinger equation \eqref{eq:SE} for this potential.
This potential allows negative-energy states when $a$ is sufficiently large.
For specific choices of $a$, all the non-negative energy eigenvalues match perfectly to that of the the harmonic oscillator and the corresponding eigenfunctions are expressed piecewise by the Hermite polynomials of different orders.
That is, the potential \eqref{eq:pot_om} with such choices of $a$ is essentially isospectral to the the harmonic oscillator.
The potential with $a=0$ clearly yields the harmonic oscillator potential.
On the other hand, for the case of $\omega=0$, the potential is also exactly solvable and describes the free electron in metals.
If one takes the limit $a\to -\infty$, the potential $V(x)$ goes to the harmonic oscillator on the positive half line.

The essential isospectrality of our potential to the harmonic oscillator implies that there exists a way of constructing potentials that are strictly isospectral to the harmonic oscillator as many as the choices of such $a$ above.
We shall discuss this construction method later in this paper.

\clearpage
The outline of this paper is as follows.
First we describe what type of problem we are to deal with in Sec.\,\ref{sec:ProblemSetting}.
The boundary conditions, including those at $x=0$, under which the Schr\"{o}dinger equation \eqref{eq:SE} is solved are introduced.
Sec.\,\ref{sec:sols} consists of two subsections; the derivation of the solution for the general case (Sec.\,\ref{sec:sol_a}) and the specific case of $a=4\ell$ ($\ell=1,2,\ldots$) (Sec.\,\ref{sec:sol_4l}) are discussed in each subsection.
In the latter, as was mentioned earlier, the Hermite-polynomial solvability arises.
Sec.\,\ref{sec:iso} constitutes the starring part of this paper, where several isospectral deformations through the Darboux transformation are computed.
We show that we can construct infinitely many potentials whose spectrum is totally identical to that of the ordinary harmonic oscillator.
Explicit examples are provided to help the readers' understand in Secs.\,\ref{sec:sol_a}, \ref{sec:sol_4l}, \ref{sec:Crum} and \ref{sec:KA}.

\section{Problem Setting}
\label{sec:ProblemSetting}
\begin{figure}
\centering
\includegraphics[scale=0.8]{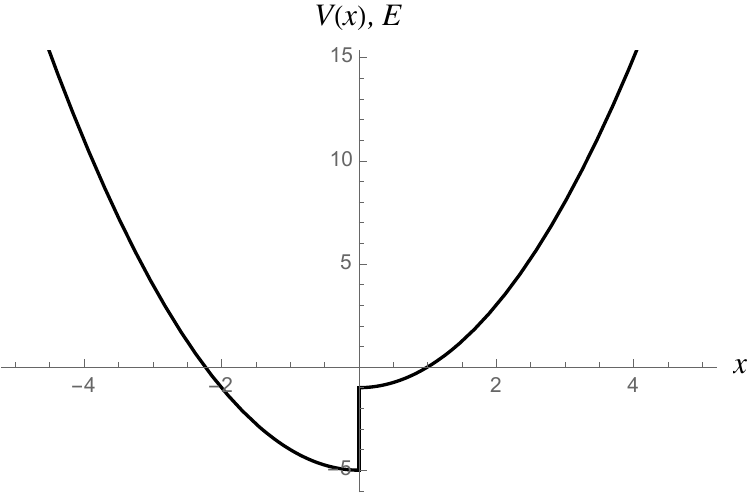}
\caption{The potential \eqref{eq:pot}.
    Here, the parameter $a$ is set to be $4$ as an example.}
\label{fig:pot}
\end{figure}

Although our potential \eqref{eq:pot_om} is not invariant under $a\leftrightarrow -a$, it is sufficient to consider the case where $a$ is a positive constant, $a>0$, by considering the parity transformation plus constant shift of the energy.
Also, one can fix $\omega=1$ without loss of generality;
\begin{equation}
V(x) = V(x;a) = \begin{cases}
	x^2 - 1 - a & (x<0) \\
	x^2 - 1 & (x>0)
\end{cases} ~,~~~
a>0 ~,
\label{eq:pot}
\end{equation}
(See Fig.\,\ref{fig:pot}).

This potential is a confining potential,
\[
\lim_{x\to +\infty} V(x) = \lim_{x\to -\infty} V(x) = +\infty ~,
\]
and has infinitely many discrete eigenvalues, $\{ E_n \}$ ($n=0,1,2,\ldots$).
The $n$-th excited state has the energy $E_n$, where $n=0$ means the ground state.
Note that this definition of the potential \eqref{eq:pot} allows eigenstates with negative energies depending on the choice of $a$.
The corresponding wavefunctions $\{ \psi_n(x) \}$ are square-integrable, $\psi_n(x) \in L^2(\mathbb{R})$,
which leads to the following boundary conditions:
\begin{equation}
\lim_{x\to +\infty}\psi(x) = \lim_{x\to -\infty}\psi(x) = 0 ~.
\label{eq:BC_inf}
\end{equation}

\subsection{The matching of wavefunctions at the origin}
In this Paper, we present the solutions of the time-independent Schr\"{o}dinger equations with the potential defined piecewise \eqref{eq:pot}.
As is the case with potentials defined piecewise, we require the continuity of the wavefunctions and their first derivatives at the origin:
\begin{equation}
\lim_{x\to 0^{+}}\psi(x) = \lim_{x\to 0^{-}}\psi(x) ~,~~~
\lim_{x\to 0^{+}}\frac{d\psi(x)}{dx} = \lim_{x\to 0^{-}}\frac{d\psi(x)}{dx} ~.
\label{eq:BC_origin}
\end{equation}

\section{The Solutions}
\label{sec:sols}

\subsection{Case for arbitrary $a$} 
\label{sec:sol_a}

Let us first consider the most general case, \textit{i.e.}, arbitrary $a>0$. 
The Schr\"{o}dinger equation:
\begin{subequations}
\begin{align}
&\left[ -\dfrac{d^2}{dx^2} + x^2 - 1 - a \right] \psi(x) = E\psi(x) & &(x<0) \label{eq:SEan} \\[1ex]
&\left[ -\dfrac{d^2}{dx^2} + x^2 - 1 \right] \psi(x) = E\psi(x) & &(x>0) ~, \label{eq:SEap}
\end{align}
\label{eq:SEa}
\end{subequations}
is piecewise solvable, whose solution is
\begin{equation}
\psi(x) =  \begin{cases}
	\mathrm{e}^{-\frac{x^2}{2}}\left[ \alpha \,{}_1F_1\left( -\dfrac{E+a}{4};\dfrac{1}{2};x^2 \right) + \beta x\, {}_1F_1\left( -\dfrac{E+a-2}{4};\dfrac{3}{2};x^2 \right)  \right] 
	& (x<0) \\[2ex]
	\mathrm{e}^{-\frac{x^2}{2}}\left[ \alpha \,{}_1F_1\left( -\dfrac{E}{4};\dfrac{1}{2};x^2 \right) + \beta x\, {}_1F_1\left( -\dfrac{E-2}{4};\dfrac{3}{2};x^2 \right)  \right] 
	& (x>0)
\end{cases} ~,
\end{equation}
with $\alpha$ and $\beta$ being constants.
The energy eigenvalues $E_n$ are determined through the boundary conditions at infinity \eqref{eq:BC_inf}, that is,
\begin{equation}
\alpha\frac{\varGamma\left( \frac{1}{2} \right)}{\varGamma\left( -\frac{E+a}{4} \right)} - \beta\frac{\varGamma\left( \frac{3}{2} \right)}{\varGamma\left( -\frac{E+a-2}{4} \right)} = 0 
~~\text{and}~~ 
\alpha\frac{\varGamma\left( \frac{1}{2} \right)}{\varGamma\left( -\frac{E}{4} \right)} + \beta\frac{\varGamma\left( \frac{3}{2} \right)}{\varGamma\left( -\frac{E-2}{4} \right)} = 0 ~.
\label{eq:E-det_a}
\end{equation}
From these equations, one obtains the following transcendental equations about $E$, whose roots are the eigenvalues:
\begin{equation}
-\frac{\varGamma\left( -\frac{E}{4} \right)}{\varGamma\left( -\frac{E-2}{4} \right)} = \frac{\varGamma\left( -\frac{E+a}{4} \right)}{\varGamma\left( -\frac{E+a-2}{4} \right)} ~.
\label{eq:E-det_a_transcental}
\end{equation}
Note however that, technically, Eq.\,\eqref{eq:E-det_a_transcental} can only be applicable to the case $a\neq 4\ell$ ($\ell=0,1,2,\ldots$), where $\alpha\neq 0$.
For the case where $a=4\ell$, we shall discuss in detail in the subsequent subsection.

\begin{figure}
\centering
\includegraphics[scale=0.9]{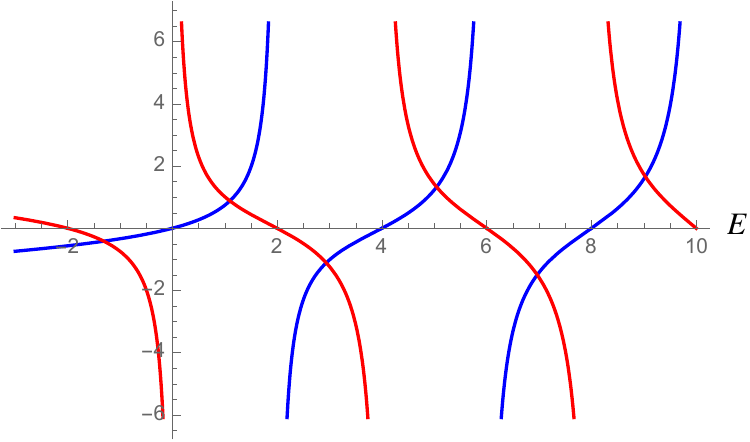}
\caption{Graphical solution of Eq.\,\eqref{eq:E-det_a_transcental}.
    The blue curves correspond to the left hand side of the equation, while the red ones are the right hand side.
    The intersections of these curves determine the energy eigenvalues.
    The numerical solutions are displayed in Tab.\,\ref{tab:ene_2}.}
\label{fig:GraphicalSol}
\end{figure}

\begin{figure}
\centering
\includegraphics[scale=0.8]{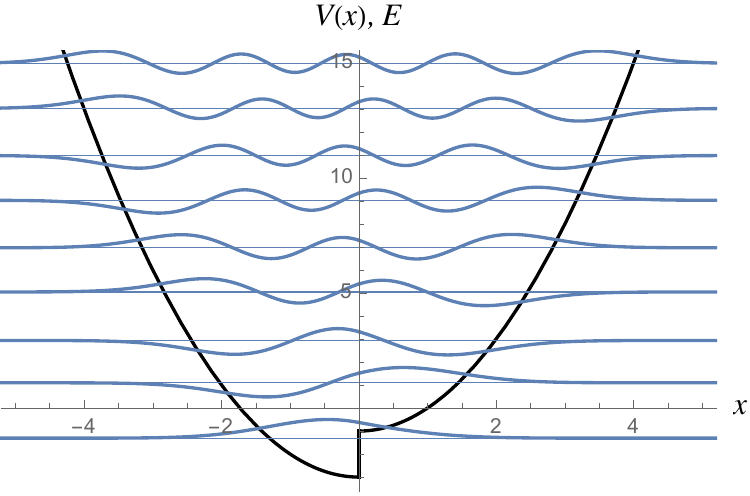}
\caption{The solutions of the eigenvalue problem \eqref{eq:SEa} with $a=2$.
    Thin blue lines show the energy spectrum, and the blue curve on each line is the corresponding eigenfunction.
    The potential $V(x;2)$ is also plotted in this figure by a black curve.}
\label{fig:sol_2}
\end{figure}

\begin{example}
We show the case of $a=2$ as an example.
We first solve Eq.\,\eqref{eq:E-det_a_transcental} with $a=2$ to obtain the energy spectrum.
This equation is transcendental, and we are to solve it graphically (See Fig.\,\ref{fig:GraphicalSol}).
The first several energy eigenvalues are displayed in Tab.\,\ref{tab:ene_2} with six digits.
The corresponding eigenfunctions are expressed in terms of $E_n$ as
\begin{equation}
\psi_n(x) \propto \begin{cases}
	\mathrm{e}^{-\frac{x^2}{2}}\left[ {}_1F_1\left( -\dfrac{E_n+2}{4};\dfrac{1}{2};x^2 \right) - 2\dfrac{\varGamma\left( -\frac{E_n-2}{4} \right)}{\varGamma\left( -\frac{E_n}{4} \right)}x\, {}_1F_1\left( -\dfrac{E_n}{4};\dfrac{3}{2};x^2 \right)  \right] 
& (x<0) \\[2.5ex]
	\mathrm{e}^{-\frac{x^2}{2}}\left[ {}_1F_1\left( -\dfrac{E_n}{4};\dfrac{1}{2};x^2 \right) - 2\dfrac{\varGamma\left( -\frac{E_n-2}{4} \right)}{\varGamma\left( -\frac{E_n}{4} \right)}x\, {}_1F_1\left( -\dfrac{E_n-2}{4};\dfrac{3}{2};x^2 \right)  \right]
& (x>0)
\end{cases} ~.
\end{equation}
The solutions are summarized in Fig.\,\ref{fig:sol_2}.
\end{example}

\begin{table}
\caption{First several energy eigenvalues for $V(x;2)$ with six digits.
    These values are obtained by solving Eq.\,\eqref{eq:E-det_a_transcental} or finding the intersections in Fig.\,\ref{fig:GraphicalSol} numerically.
    The energy gaps between two successive eigenstates are roughly $2$, while those for the harmonic oscillator are exactly $2$ in our units.}
\label{tab:ene_2}
\centering
\begin{tabular}{cr}
\toprule
$n$ & $E_n$~~~ \\
\midrule
$0$ & $-1.30908$ \\
$1$ & $1.09714$ \\
$2$ & $2.93715$ \\
$3$ & $5.04459$ \\
$4$ & $6.96479$ \\
$5$ & $9.02870$ \\
$6$ & $10.9756$ \\
\bottomrule
\end{tabular}
\end{table}

\subsection{Hermite-polynomial solutions for $a=4\ell$ ($\ell = 1,2,\ldots$)} 
\label{sec:sol_4l}

Next let us consider the case where $a$ is a multiple of $4$, $a=4\ell$ or $V(x;a) = V(x;4\ell)$ ($\ell=1,2,\ldots$), where all the non-negative eigenfunctions are expressed with the Hermite polynomials as we shall demonstrate bellow.
The Schr\"{o}dinger equation is
\begin{subequations}
\begin{align}
&\left[ -\dfrac{d^2}{dx^2} + x^2 - 1 - 4\ell \right] \psi(x) = E\psi(x) & &(x<0) \label{eq:SE4ln} \\[1ex]
&\left[ -\dfrac{d^2}{dx^2} + x^2 - 1 \right] \psi(x) = E\psi(x) & &(x>0) ~. \label{eq:SE4lp}
\end{align}
\label{eq:SE4l}
\end{subequations}
For this case, the transcendental equations \eqref{eq:E-det_a} reduce to rather simple algebraic equations, and all the eigenfunctions with $E\geqslant 0$ are expressed by the Hermite polynomials $\{H_n(x)\}$.

Before we solve Eq.\,\eqref{eq:SE4l}, let us summarize the solutions of Eqs.\,\eqref{eq:SE4ln} and \eqref{eq:SE4lp} solved on the real line, $x\in\mathbb{R}$, respectively.
They are
\begin{subequations}
\begin{align}
&\eqref{eq:SE4ln}\Rightarrow &
&E_n^{(-)} = 2n-4\ell ~,~~~ \psi_n^{(-)}(x) = \mathcal{N}_n^{(-)}\mathrm{e}^{-\frac{x^2}{2}}H_n(x) ~,~~~
n = 0,1,2,\ldots ~, \qquad~ \label{eq:soln_SE4l_E+} \\
&\eqref{eq:SE4lp}\Rightarrow &
&E_n^{(+)} = 2n ~,~~~ \psi_n^{(+)}(x) = \mathcal{N}_n^{(+)}\mathrm{e}^{-\frac{x^2}{2}}H_n(x) ~,~~~
n = 0,1,2,\ldots ~. \qquad ~ \label{eq:solp_SE4l_E+}
\end{align}
\label{eq:sol_SE4l_E+}
\end{subequations}
Here, they share all non-negative energy eigenvalues.
This suggests that $E_j=2j$ ($j=0,1,2,\ldots$) are the eigenvalues of the system \eqref{eq:SE4l} if the coefficients $\mathcal{N}_n^{(\pm)}$ are chosen so that they satisfy the continuity conditions of the wavefunction and its first derivative at the origin \eqref{eq:BC_origin}.

For $E=0$, the wavefunctions are 
\begin{equation}
\psi_{2\ell}^{(-)}(x) \propto \mathrm{e}^{-\frac{x^2}{2}}H_{2\ell}(x) ~,~~~
\psi_0^{(+)}(x) \propto \mathrm{e}^{-\frac{x^2}{2}} ~.
\end{equation}
Since
\begin{equation}
\mathrm{e}^{-\frac{0^2}{2}}H_{2\ell}(0) = (-1)^{\ell}\frac{(2\ell)!}{\ell!} ~,~~~
\mathrm{e}^{-\frac{0^2}{2}} = 1 ~,~~~
\frac{d\psi_{2\ell}^{(-)}(0)}{dx} = \frac{d\psi_0^{(+)}(0)}{dx} = 0 ~,
\end{equation}
the continuity conditions at the origin \eqref{eq:BC_origin} yields the wavefunction with $E=0$ for the original problem \eqref{eq:SE4l} in the following form:
\begin{equation}
\psi_{E=0}(x) = \begin{cases}
	(-1)^{\ell}\dfrac{\ell!}{(2\ell)!}\mathrm{e}^{-\frac{x^2}{2}}H_{2\ell}(x) & (x<0) \\
	\mathrm{e}^{-\frac{x^2}{2}} & (x>0)
\end{cases} ~.
\end{equation}
This is a square-integrable, smooth function of $\ell$ nodes, so from the oscillation theorem, one can safely say that this corresponds to the $\ell$-th excited state, $\psi_{E=0}(x)\equiv\psi_{\ell}(x)$.

Similarly, one can obtain the square-integrable wavefunctions for all the non-negative energies.
The $n$-th excited-state energy eigenvalue for Eq.\,\eqref{eq:SE4l} is $E_n=2(n-\ell)$, where $n$ is greater than or equal to $\ell$, and the corresponding eigenfunction is
\begin{equation}
\psi_n(x) = \begin{cases}
	\mathcal{N}_{n}\,\mathrm{e}^{-\frac{x^2}{2}}H_{n+\ell}(x) & (x<0) \\
	\mathrm{e}^{-\frac{x^2}{2}}H_{n-\ell}(x) & (x>0) 
\end{cases} ~,~~~
n = \ell,\ell+1,\ell+2,\ldots ~,
\end{equation}
where
\begin{subequations}
\begin{align}
\mathcal{N}_{n}
&= (-1)^{\ell}\frac{(n-\ell)!\left( \frac{n+\ell}{2} \right)!}{\left( n+\ell \right)!\left( \frac{n-\ell}{2} \right)!} 
&&\text{if $(n-\ell)$ is even} ~, \\
\mathcal{N}_{n}
&= (-1)^{\ell}\frac{(n-\ell)!\left( \frac{n+\ell-1}{2} \right)!}{\left( n+\ell \right)!\left( \frac{n-\ell-1}{2} \right)!} 
&&\text{if $(n-\ell)$ is odd} ~.
\end{align}
\end{subequations}
Note that this wavefunction satisfies either Neumann or Dirichlet boundary condition at the origin.
For the case where $(n-\ell)$ is even, it complies with
\begin{equation}
\psi_{n}(0) \neq 0 ~,~~~
\frac{d\psi_{n}(0)}{dx} = 0
\qquad\text{: {\it Neumann} boundary condition,}
\end{equation}
while for odd $(n-\ell)$,
\begin{equation}
\psi_{n}(0) =  0 ~,~~~
\frac{d\psi_{n}(0)}{dx} \neq 0
\qquad\text{: {\it Dirichlet} boundary condition.}
\end{equation}
For $E \geqslant 0$, our problem \eqref{eq:SE4l} would be rephrased as finding square-integrable solutions of \eqref{eq:SEa} under the Neumann/Dirichlet boundary condition at the origin.

For the $\ell$ lowest eigenstates, the eigenfunctions are not expressed by the Hermite polynomials anymore. 
So we need to go back to a problem of solving the piecewise differential equation \eqref{eq:SE4l} under the boundary conditions \eqref{eq:BC_inf},\eqref{eq:BC_origin}.
Eq.\,\eqref{eq:BC_inf} yields 
\begin{equation}
-\frac{\varGamma\left( -\frac{E}{4} \right)}{\varGamma\left( -\frac{E-2}{4} \right)} = \frac{\varGamma\left( -\frac{E+4\ell}{4} \right)}{\varGamma\left( -\frac{E+4\ell-2}{4} \right)} ~.
\end{equation}
This may look a transcendental equation, however, after some algebras,  one finds that it is a degree $\ell$ algebraic equation:
\begin{equation}
-\prod_{k=1}^{\ell}(E+4k-2) = \prod_{k=1}^{\ell}(E+4k) ~.
\label{eq:algebraic}
\end{equation}
Here the roots of this equation are denoted by $E_0,E_1,E_2,\ldots, E_{\ell-1}$ from the lowest to higher.
Therefore, for $E<0$, the wavefunctions are 
\begin{multline}
\psi_n(x) = \begin{cases}
	\displaystyle
	\mathrm{e}^{-\frac{x^2}{2}}\left[ {}_1F_1\left( -\frac{E_n+4\ell}{4};\frac{1}{2};x^2 \right) - 2\frac{\varGamma\left( -\frac{E_n-2}{4} \right)}{\varGamma\left( -\frac{E_n}{4} \right)}x\, {}_1F_1\left( -\frac{E_n+4\ell-2}{4};\frac{3}{2};x^2 \right)  \right] & (x<0) \\[2.5ex]
	\displaystyle
	\mathrm{e}^{-\frac{x^2}{2}}\left[ {}_1F_1\left( -\frac{E_n}{4};\frac{1}{2};x^2 \right) - 2\frac{\varGamma\left( -\frac{E_n-2}{4} \right)}{\varGamma\left( -\frac{E_n}{4} \right)}x\, {}_1F_1\left( -\frac{E_n-2}{4};\frac{3}{2};x^2 \right)  \right] & (x>0) 
\end{cases} ~, \\
n=0,1,\ldots,\ell-1 ~.
\end{multline}

\begin{figure}[t]
\centering
\includegraphics[scale=0.8]{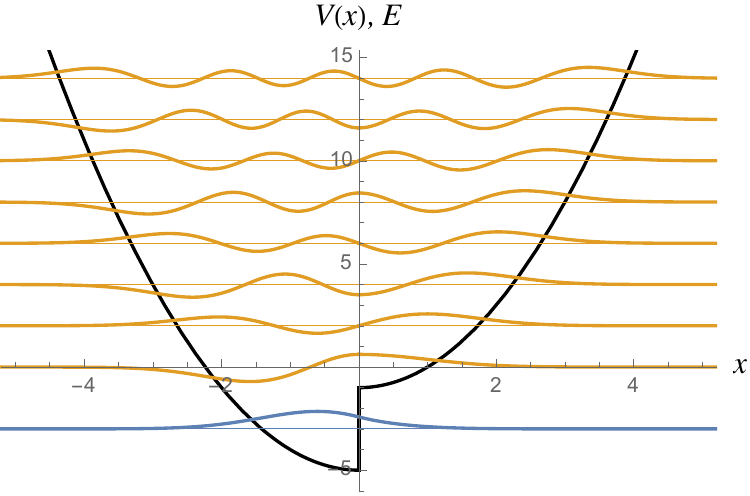}
\caption{The solutions of the eigenvalue problem \eqref{eq:SE4l} with $\ell=1$.
    The potential $V(x;4)$ is displayed in this figure by a black curve.
    Thin lines show the energy spectrum, and the colored curve on each line is the corresponding eigenfunction.
    The states plotted in yellow possess the Hermite-polynomial solvability, while that colored in blue does not.}
\label{fig:sol_4}
\end{figure}

\begin{example}
The energy eigenvalues and the corresponding eigenfunctions for $\ell=1$ are
\begin{align}
E_0 &= -3 ~,~~~
\psi_0(x) = \begin{cases}
	\displaystyle
	\mathrm{e}^{-\frac{x^2}{2}}\left[ {}_1F_1\left( -\frac{1}{4};\frac{1}{2};x^2 \right) - 2\frac{\varGamma\left( \frac{5}{4} \right)}{\varGamma\left( \frac{3}{4} \right)}x\, {}_1F_1\left( \frac{1}{4};\frac{3}{2};x^2 \right)  \right] & (x<0)  \\[2.5ex]
	\displaystyle
	\mathrm{e}^{-\frac{x^2}{2}}\left[ {}_1F_1\left( \frac{3}{4};\frac{1}{2};x^2 \right) - 2\frac{\varGamma\left( \frac{5}{4} \right)}{\varGamma\left( \frac{3}{4} \right)}x\, {}_1F_1\left( \frac{5}{4};\frac{3}{2};x^2 \right)  \right] & (x>0)
\end{cases} ~, \\
E_n &= 2(n-1) ~,~~~
\psi_n(x) = \begin{cases}
	\mathcal{N}_{n}\,\mathrm{e}^{-\frac{x^2}{2}}H_{n+1}(x) & (x<0) \\
	\mathrm{e}^{-\frac{x^2}{2}}H_{n-1}(x) & (x>0) 
\end{cases} ~,~~~ 
n = 1,2,3,\ldots ~,
\end{align}
where
\begin{subequations}
\begin{align}
\mathcal{N}_{n}
&= -\frac{1}{2n} 
&&\text{if $n$ is odd} ~, \\
\mathcal{N}_{n}
&= -\frac{1}{2(n+1)}
&&\text{if $n$ is even} ~.
\end{align}
\end{subequations}
They are summarized in Fig.\,\ref{fig:sol_4}.
\end{example}

\begin{table}[]
\caption{The negative energy eigenvalues for $V(x;24)$ with six digits.
    These values are obtained by solving Eq.\,\eqref{eq:algebraic} with $\ell=6$.
    The explicit forms of the roots are shown in Appendix~\ref{sec:ene_l6}.
    The energy gaps between two successive eigenstates are roughly $4$, while those for the harmonic oscillator on the positive half line are exactly $4$ in our units.}
\label{tab:ene_l6}
\centering
\medskip
\begin{tabular}{cr}
\toprule
$n$ & $E_n$~~~ \\
\midrule
$0$ & $-22.4357$ \\
$1$ & $-18.6885$ \\
$2$ & $-14.8995$ \\
$3$ & $-11.1005$ \\
$4$ & $-7.31152$ \\
$5$ & $-3.56427$ \\
\bottomrule
\end{tabular}
\end{table}

\begin{figure}[t]
\centering
\includegraphics[scale=0.8]{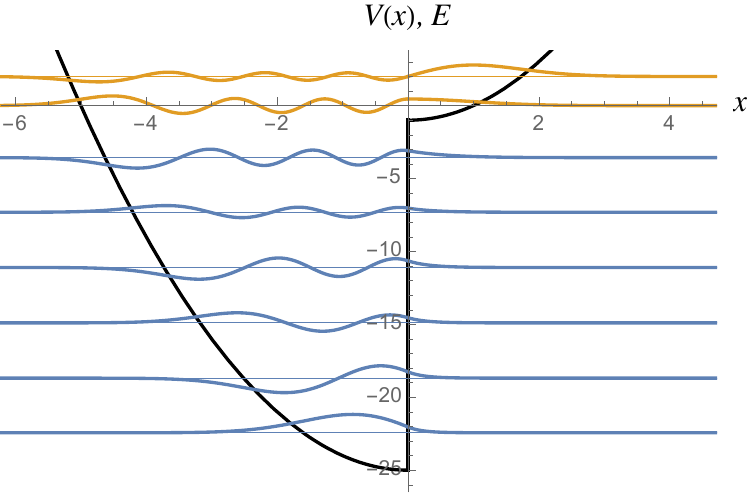}
\caption{The solutions of the eigenvalue problem \eqref{eq:SE4l} with $\ell=6$.
    The potential $V(x;24)$ is displayed in this figure by a black curve.
    Thin lines show the energy spectrum, and the colored curve on each line is the corresponding eigenfunction.
    The states plotted in yellow possess the Hermite-polynomial solvability, while those colored in blue do not.}
\label{fig:sol_24}
\end{figure}

\begin{example}
\label{eg:l6}
The energy eigenvalues and the corresponding eigenfunctions for $\ell=6$ are
\begin{align}
\psi_n(x) &= \begin{cases}
	\displaystyle
	\mathrm{e}^{-\frac{x^2}{2}}\left[ {}_1F_1\left( -\frac{E_n+24}{4};\frac{1}{2};x^2 \right) - 2\frac{\varGamma\left( -\frac{E_n-2}{4} \right)}{\varGamma\left( -\frac{E_n}{4} \right)}x\, {}_1F_1\left( -\frac{E_n+22}{4};\frac{3}{2};x^2 \right)  \right] & (x<0) \\[2.5ex]
	\displaystyle
	\mathrm{e}^{-\frac{x^2}{2}}\left[ {}_1F_1\left( -\frac{E_n}{4};\frac{1}{2};x^2 \right) - 2\frac{\varGamma\left( -\frac{E_n-2}{4} \right)}{\varGamma\left( -\frac{E_n}{4} \right)}x\, {}_1F_1\left( -\frac{E_n-2}{4};\frac{3}{2};x^2 \right)  \right] & (x>0) 
\end{cases} ~, \nonumber \\
&\hspace{0.7\textwidth} n=0,1,\ldots, 5 \\
E_n &= 2(n-6) ~,~~~
\psi_n(x) = \begin{cases}
	\mathcal{N}_{n}\,\mathrm{e}^{-\frac{x^2}{2}}H_{n+6}(x) & (x<0) \\
	\mathrm{e}^{-\frac{x^2}{2}}H_{n-6}(x) & (x>0) 
\end{cases} ~,~~~ 
n = 6,7,8,\ldots ~,
\end{align}
where
\begin{subequations}
\begin{align}
\mathcal{N}_{n}
&= \frac{(n-6)!\left( \frac{n+6}{2} \right)!}{\left( n+6 \right)!\left( \frac{n-6}{2} \right)!} 
&&\text{if $n$ is even} ~, \\
\mathcal{N}_{n}
&= \frac{(n-6)!\left( \frac{n+5}{2} \right)!}{\left( n+6 \right)!\left( \frac{n-7}{2} \right)!} 
&&\text{if $n$ is odd} ~.
\end{align}
\end{subequations}
The negative energy eigenvalues are displayed in Tab.\,\ref{tab:ene_l6} (Also see Appendix~\ref{sec:ene_l6} for their explicit forms).
They are summarized in Fig.\,\ref{fig:sol_24}.
\end{example}

\subsubsection{On the algebraic equation \eqref{eq:algebraic}}
Although we do not have a general formula for solving the $\ell$-th degree algebraic equation \eqref{eq:algebraic}, it turns out that the $\ell$ roots have the following property: \textit{if $E=-1-2\ell+\alpha$ is a root of Eq.\,\eqref{eq:algebraic}, $E=-1-2\ell-\alpha$ is also a root.}
Here, we give the proof:

\paragraph*{Proof.}
We change the variable $E\to \tilde{E}-1-2\ell$.
By this, the equation \eqref{eq:algebraic} becomes
\begin{equation}
\prod_{k=1}^{\ell}(\tilde{E}+4k-2\ell-3) + \prod_{k=1}^{\ell}(\tilde{E}+4k-2\ell-1) = 0 ~.
\label{eq:algebraic2}
\end{equation}
Now under the transformation: $\tilde{E}\to -\tilde{E}$, the equation transforms
\begin{align*}
&\prod_{k=1}^{\ell}(-\widetilde{E}+4k-2\ell-3) + \prod_{k=1}^{\ell}(-\widetilde{E}+4k-2\ell-1) \\
=\,&(-1)^{\ell}\left( \prod_{k=1}^{\ell}(\widetilde{E}-4k+2\ell+3) + \prod_{k=1}^{\ell}(\widetilde{E}-4k+2\ell+1) \right) \\
=\,&(-1)^{\ell}\left( \prod_{k=1}^{\ell}(\widetilde{E}+4k-2\ell-3) + \prod_{k=1}^{\ell}(\widetilde{E}+4k-2\ell-1) \right) 
= (-1)^{\ell} \times \text{[l.h.s. of Eq.\,\eqref{eq:algebraic2}]} ~.
\end{align*}
Therefore, if $\tilde{E}=\alpha$ satisfies Eq.\,\eqref{eq:algebraic2}, $\tilde{E}=-\alpha$ also satisfies the equation. \hfill{$\square$}

\bigskip
Note that this property guarantees even numbers of roots of the algebraic equation.
For odd-$\ell$ case, $E=-1-2\ell$ is also a root of Eq.\,\eqref{eq:algebraic}, which corresponds to $\alpha = 0$ above.

\section{Isospectral Hamiltonians}
\label{sec:iso}

\subsection{Crum's theorem}
\label{sec:Crum}

Let $\mathcal{H}_{\ell}^{[0]}$ denote the Hamiltonian of our one-dimensional quantum mechanical system with $a=4\ell$ \eqref{eq:SE4l}:
\begin{equation}
\mathcal{H}_{\ell}^{[0]} = -\frac{d^2}{dx^2} + V(x;4\ell) ~.
\end{equation}
Also, we write $E_{\ell,n}$ and $\psi_{\ell,n}^{[0]}$ ($n=0,1,2,\ldots$) for the energy eigenvalues and the corresponding eigenfunctions hereafter.

According to the Crum's theorem~\cite{10.1093/qmath/6.1.121}, there are infinitely many associated Hamiltonian systems $\mathcal{H}_{\ell}^{[M]}, M=1,2,\ldots$, which are essentially isospectral to $\mathcal{H}_{\ell}^{[0]}$.
They are
\begin{equation}
\mathcal{H}_{\ell}^{[M]} \coloneqq 
\mathcal{H}_{\ell}^{[0]} - 2\frac{d^2}{dx^2}\ln\mathrm{W}\left[ \psi_{\ell,0}^{[0]},\psi_{\ell,1}^{[0]},\psi_{\ell,2}^{[0]},\ldots,\psi_{\ell,M-1}^{[0]} \right](x) ~,
\label{eq:Ham_DC}
\end{equation}
in which $\mathrm{W}[f_1,\ldots,f_m](x)$ is the Wronskian defined as 
\begin{equation}
\mathrm{W}[f_1,\ldots,f_m](x) \coloneqq \det\left( \frac{d^{j-1}f_k(x)}{dx^{j-1}} \right)_{1\leqslant j,k\leqslant m} ~.
\end{equation}
$\mathcal{H}_{\ell}^{[M]}$ is such a system that the $M$ lowest eigenstates are deleted from $\mathcal{H}_{\ell}^{[0]}$, and shares all the eigenvalues above $E_{\ell,M}$ with $\mathcal{H}_{\ell}^{[0]}$.
The corresponding eigenfunctions $\{ \psi_{\ell,n}^{[M]} \}$ are related to $\{ \psi_{\ell,n}^{[0]} \}$ by the Dourboux--Crum transformation~\cite{darboux,Bagrov:1995aa}:
\begin{equation}
\psi_{\ell,n}^{[M]}(x) =
\frac{\mathrm{W}\left[ \psi_{\ell,0}^{[0]},\psi_{\ell,1}^{[0]},\ldots,\psi_{\ell,M-1}^{[0]},\psi_{\ell,n+M}^{[0]} \right](x)}{\mathrm{W}\left[ \psi_{\ell,0}^{[0]},\psi_{\ell,1}^{[0]},\ldots,\psi_{\ell,M-1}^{[0]} \right](x)} ~,
\end{equation}
which satisfies the following Schr\"{o}dinger equation: 
\begin{equation}
\mathcal{H}_{\ell}^{[M]}\psi_{\ell,n}^{[M]}(x) = E_{\ell,n+M}\psi_{\ell,n}^{[M]}(x) ~,~~~
n=0,1,2,\ldots ~.
\end{equation}

\begin{figure}[p]
\centering
\includegraphics[scale=0.8]{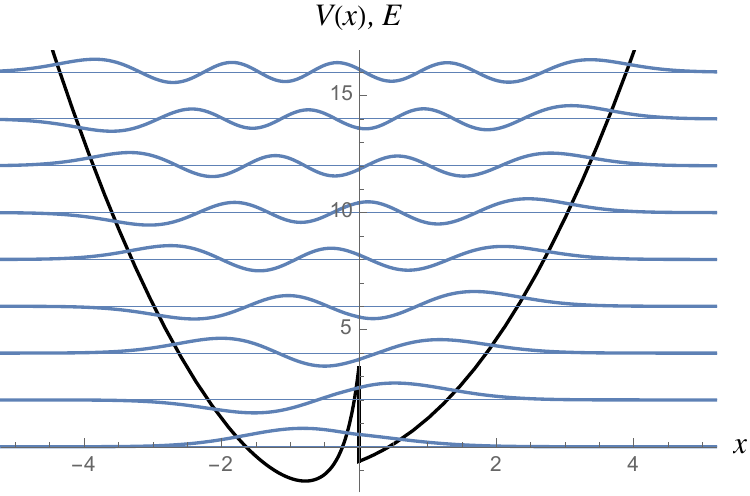}
\caption{The solutions of the eigenvalue problem \eqref{eq:Ham_DC} with $\ell=1$ and $M=1$.
    Thin blue lines show the energy spectrum, and the blue curve on each line is the corresponding eigenfunction.
    The potential of this system is also plotted in this figure by a black curve.}
\label{fig:l1M1}
\end{figure}

\begin{example}
Let us take $\ell=1$ and $M=1$ as an example.
In the context of supersymmetric quantum mechanics~\cite{COOPER1995267}, the associated Hamiltonian with $M=1$ is often referred to as the SUSY partner.
Here, the Hamiltonian is 
\begin{equation}
\mathcal{H}_{1}^{[1]} = \mathcal{H}_{1}^{[0]} - 2\frac{d^2}{dx^2}\ln\psi_{1,0}^{[0]}(x) ~,~~~
\mathcal{H}_{1}^{[1]}\psi_{1,n}^{[1]}(x) = E_{1,n+1}\psi_{1,n}^{[1]}(x) ~,~~~
n=0,1,2,\ldots ~,
\end{equation}
where
\begin{equation}
\psi_{1,n}^{[1]}(x) =
\frac{\mathrm{W}\left[ \psi_{1,0}^{[0]},\psi_{1,n+1}^{[0]} \right](x)}{\psi_{1,0}^{[0]}(x)} ~.
\end{equation}
We plot them for the first several eigenstates in Fig.\,\ref{fig:l1M1}.

Note from the explicit calculation that $\mathcal{H}_{1}^{[0]}$ and $\mathcal{H}_{1}^{[1]}$ are not shape invariant~\cite{Gendenshtein:1983skv}.
The same can be applied to $\mathcal{H}_{\ell}^{[M]}$ and $\mathcal{H}_{\ell}^{[M+1]}$.
One might guess that the Hermite-polynomial solvability of our system is due to the shape invariance as in the case of the harmonic oscillator.
However, this is not the case.
\end{example}

\subsection{Isospectrality to 1-d harmonic oscillator}
\begin{figure}[p]
\centering
\includegraphics[scale=0.8]{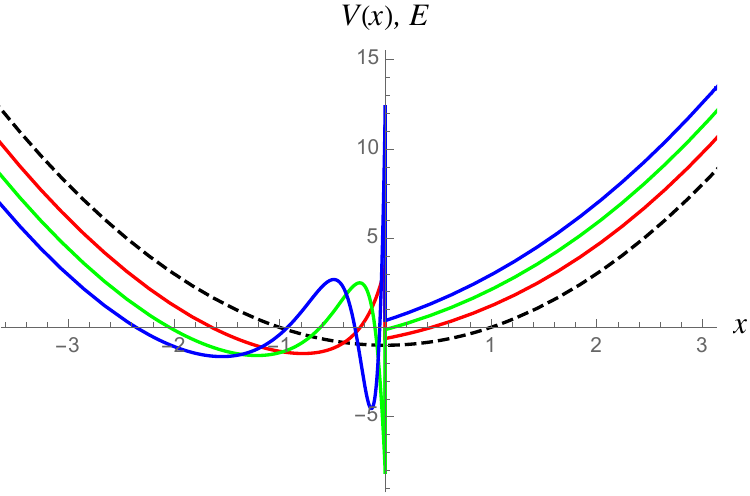}
\caption{The sequence $\{ \mathcal{H}_{\ell}^{[\ell]} \}$. 
    The potentials for $\ell=1,2,3$ are plotted in red, green and blue respectively.
    Those potentials are all isospectral to the 1-d harmonic oscillator potential (dashed black curve).
    $\mathcal{H}_{\ell}^{[\ell]}$'s are non-analytic at $x=0$, but never diverge.}
\label{fig:sequence}
\end{figure}

We choose $M=\ell$ here.
The resulting Hamiltonian is 
\begin{equation}
\mathcal{H}_{\ell}^{[\ell]} =
\mathcal{H}_{\ell}^{[0]} - 2\frac{d^2}{dx^2}\ln\mathrm{W}\left[ \psi_{\ell,0}^{[0]},\psi_{\ell,1}^{[0]},\psi_{\ell,2}^{[0]},\ldots,\psi_{\ell,\ell-1}^{[0]} \right](x) ~,
\label{eq:ham_iso}
\end{equation}
which corresponds to the deletion of all negative-energy states of $\mathcal{H}_{\ell}^{[0]}$.
This Hamiltonian is strictly isospectral to the 1-d harmonic oscillator potential $\mathcal{H}_{\rm HO}(x)=x^2-1$.
Since $\ell$ can be any positive integer, we now have infinitely many isospectral potentials of the harmonic oscillator in our procedure above.
We plot the first several potentials of the sequence $\{ \mathcal{H}_{\ell}^{[\ell]} \}$ ($\ell = 1,2,\ldots$) in Fig.\,\ref{fig:sequence}.

\subsection{Further deformation: Krein--Adler transformation}
\label{sec:KA}

A further generalization of the isospectral deformation in Sec.\,\ref{sec:Crum}, \textit{i.e.}, deletions of the eigenstates from the original system, was formulated by Krein and Adler independently~\cite{Kre57,Adler:1994aa}.
During this deformation, the eigenstates with the following indices are deleted:
\begin{equation}
\mathcal{D} \coloneqq \{ d_1,d_1+1 < d_2,d_2+1 < \cdots < d_N,d_N+1 \} ~,~~~
d_1,\ldots,d_N \in \mathbb{Z}_{\geqslant 0} ~,
\end{equation}
where taking $d_1=0$ and $d_{j+1}=d_j+2$ for all $j$ corresponds to the case of the Crum's theorem.
The Hamiltonian is 
\begin{equation}
\mathcal{H}_{\ell}^{\mathcal{D}} \coloneqq 
\mathcal{H}_{\ell}^{[0]} - 2\frac{d^2}{dx^2}\ln\mathrm{W}\left[ \psi_{\ell,d_1}^{[0]},\psi_{\ell,d_1+1}^{[0]},\ldots,\psi_{\ell,d_N+1}^{[0]} \right](x) ~,
\label{eq:Ham_KA}
\end{equation}
which shares all the energy spectrum with $\mathcal{H}_{\ell}^{[0]}$ except that those indexed by $\mathcal{D}$ are deleted.
The eigenfunctions are
\begin{equation}
\psi_{\ell,\tilde{n}}^{\mathcal{D}}(x) \coloneqq 
\frac{\mathrm{W}\left[ \psi_{\ell,d_1}^{[0]},\psi_{\ell,d_1+1}^{[0]},\ldots,\psi_{\ell,d_N+1}^{[0]},\psi_{\ell,\tilde{n}}^{[0]} \right](x)}{\mathrm{W}\left[  \psi_{\ell,d_1}^{[0]},\psi_{\ell,d_1+1}^{[0]},\ldots,\psi_{\ell,d_N+1}^{[0]} \right](x)} ~,~~~
\tilde{n} \in \mathbb{Z}_{\geqslant 0} \backslash \mathcal{D} ~,
\end{equation}
satisfying
\begin{equation}
\mathcal{H}_{\ell}^{\mathcal{D}}\psi_{\ell,\tilde{n}}^{\mathcal{D}}(x) = E_{\ell,\tilde{n}}\psi_{\ell,\tilde{n}}^{\mathcal{D}}(x) ~.
\end{equation}

\begin{figure}[t]
\vspace{-1.5ex}
\centering
\includegraphics[scale=0.8]{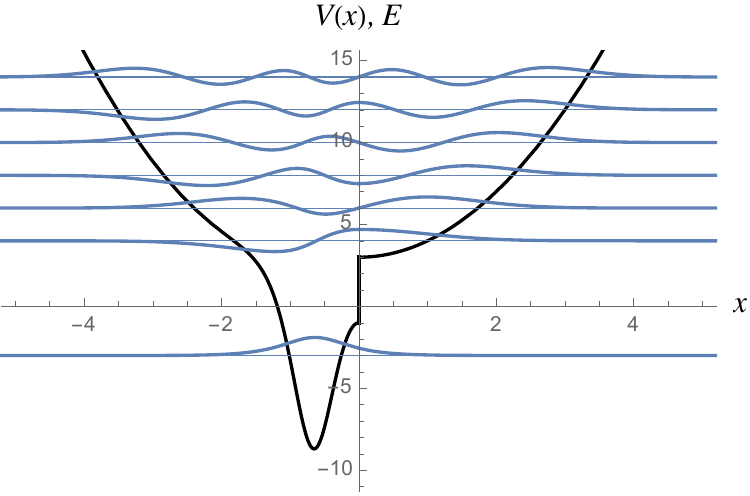}
\caption{The solutions of the eigenvalue problem \eqref{eq:Ham_KA} with $\ell=1$ and $\mathcal{D}=\{ 1,2 \}$.
    Thin blue lines show the energy spectrum, and the blue curve on each line is the corresponding eigenfunction.
    The potential of this system is also plotted in this figure by a black curve.}
\label{fig:KA_12}
\end{figure}

\begin{example}
Taking $\ell=1$ and $\mathcal{D}=\{ 1,2 \}$, we get
\begin{equation}
\mathcal{H}_{1}^{\{ 1,2 \}} = \mathcal{H}_{1}^{[0]} - 2\frac{d^2}{dx^2}\ln\mathrm{W}\left[ \psi_{1,1}^{[0]},\psi_{1,2}^{[0]} \right](x) ~,~~~
\mathcal{H}_{1}^{\{ 1,2 \}}\psi_{1,\tilde{n}}^{\{ 1,2 \}}(x) = E_{1,\tilde{n}}\psi_{1,\tilde{n}}^{\{ 1,2 \}}(x) ~,~~~
\tilde{n}=0,3,4,\ldots ~,
\end{equation}
where
\begin{equation}
\psi_{1,\tilde{n}}^{\{ 1,2 \}}(x) =
\frac{\mathrm{W}\left[ \psi_{1,1}^{[0]},\psi_{1,2}^{[0]},\psi_{\ell,\tilde{n}}^{[0]} \right](x)}{\mathrm{W}\left[ \psi_{1,1}^{[0]},\psi_{1,2}^{[0]} \right](x)} ~,
\end{equation}
(See Fig.\,\ref{fig:KA_12}).
\end{example}

\section{Conclusion}
In this paper, we have presented analytical solutions of the static Schr\"{o}dinger equation with a potential \eqref{eq:pot}, and constructed several essentially isospectral, exactly solvable potentials through the Darboux transformations.
The potential \eqref{eq:pot} contains a singularity at $x=0$.
As in the case of other potentials defined by piecewise analytic functions, we have employed the matching of wavefunctions to solve the problem.
The newly constructed potentials through the Darboux transformations \eqref{eq:Ham_DC}, \eqref{eq:Ham_KA} are also analytic except at the origin, succeeding the singular nature of our original potential \eqref{eq:pot}.

The eigenvalue problem of the Hamiltonian with the potential \eqref{eq:pot} is in general solved by using confluent hypergeometric functions, and no eigenvalues are typically identical to those of the ordinary harmonic oscillator.
On the other hand, our computation shows that when we take $a=4\ell$ ($\ell = 1,2,\ldots$), the Hermite-polynomial solvability arises for the non-negative energy states.
The main part of the wavefunction is expressed by the Hermite polynomial.
Also, those states are isospectral to the harmonic oscillator.
One might think that this solvability is guaranteed by the same mathematical structure as the harmonic oscillator, \textit{i.e.}, shape invariance, but we have given a negative answer to this point.

By eliminating all negative-energy states, which are neither Hermite-polynomially solvable states nor isospectral to the harmonic oscillator, we have obtained a strictly isospectral potential to the harmonic oscillator.
Since the parameter $\ell$ can be any positive integer, we can obtain infinitely many potentials that are strictly isospectral to the harmonic oscillator by this procedure.
The elimination of all negative-energy states is realized by multiple applications of the Darboux transformation.
The Darboux transformation allows many other deformations of the Hamiltonian, where all of the resulting deformed Hamiltonians are essentially isospectral to the harmonic oscillator.

\appendix
\section{Explicit forms of the negative energy eigenvalues for $V(x;24)$}
\label{sec:ene_l6}

We list the explicit forms of the negative energy eigenvalues in Example~\ref{eg:l6}.
In the following, $\mathrm{i}=\sqrt{-1}$ as usual.
\begin{align}
E_0 &= -13-\sqrt{\frac{1}{3} \left(2 \sqrt[3]{28315+\mathrm{i}\,216 \sqrt{43798}}+\frac{2834}{\sqrt[3]{28315+\mathrm{i}\,216 \sqrt{43798}}}+125\right)} \nonumber \\
&\approx -22.4357 ~, \\
E_1 &= -13-\sqrt{\frac{1}{3} \left(\left(-1-\mathrm{i} \sqrt{3}\right) \sqrt[3]{28315+\mathrm{i}\,216 \sqrt{43798}}+\frac{1417 \left(-1 + \mathrm{i}\sqrt{3}\right)}{\sqrt[3]{28315+\mathrm{i}\,216 \sqrt{43798}}}+125\right)} \nonumber \\
&\approx -18.6885 ~, \\
E_2 &= -13-\sqrt{\frac{1}{3} \left( \left(-1+\mathrm{i}\sqrt{3}\right) \sqrt[3]{28315+\mathrm{i}\,216 \sqrt{43798}}-\frac{1417 \left(1+\mathrm{i} \sqrt{3}\right)}{\sqrt[3]{28315+\mathrm{i}\,216 \sqrt{43798}}}+125\right)} \nonumber \\
&\approx -14.8995 ~, \\
E_3 &= -13+\sqrt{\frac{1}{3} \left( \left(-1+\mathrm{i}\sqrt{3}\right) \sqrt[3]{28315+\mathrm{i}\,216 \sqrt{43798}}-\frac{1417 \left(1+\mathrm{i} \sqrt{3}\right)}{\sqrt[3]{28315+\mathrm{i}\,216 \sqrt{43798}}}+125\right)} \nonumber \\
&\approx -11.1005 ~, \\
E_4 &= -13+\sqrt{\frac{1}{3} \left(\left(-1-\mathrm{i} \sqrt{3}\right) \sqrt[3]{28315+\mathrm{i}\,216 \sqrt{43798}}+\frac{1417 \left(-1 + \mathrm{i}\sqrt{3}\right)}{\sqrt[3]{28315+\mathrm{i}\,216 \sqrt{43798}}}+125\right)} \nonumber \\
&\approx -7.31152 ~, \\
E_5 &= -13+\sqrt{\frac{1}{3} \left(2 \sqrt[3]{28315+\mathrm{i}\,216 \sqrt{43798}}+\frac{2834}{\sqrt[3]{28315+\mathrm{i}\,216 \sqrt{43798}}}+125\right)} \nonumber \\
&\approx -3.56427 ~.
\end{align}

\bigskip
\section*{Acknowledgment}
The authors would like to thank Ryu Sasaki for his careful and useful advice and comments.
We also appreciate Luiz Agostinho Ferreira for his kind hospitality in Instituto de F\'{i}sica de S\~{a}o Carlos of Universidade de S\~{a}o Paulo (IFSC/USP).
YN is supported by JST SPRING Grant Number JPMJSP2151 and was supported by the Sasakawa Scientific Research Grant from The Japan Science Society (No. 2022-2011).
NS is supported in part by JSPS KAKENHI Grant Number JP B20K03278.

\bibliographystyle{naturemag}
\bibliography{Dislocation_HO}

\end{document}